\documentclass[aps, prd, twocolumn, lengthcheck, superscriptaddress, 
nofootinbib]{revtex4-1}

\usepackage{amsmath,amssymb,amsfonts}
\usepackage{slashed}
\usepackage{graphicx}
\usepackage{dcolumn}
\usepackage{bm}
\usepackage{tikz}

\def\bi{\bibitem}

\def\la{\langle}\def\ra{\rangle}
\def\be{\begin{eqnarray}}\def\ee{\end{eqnarray}}
\def\lsim{\mathrel{\rlap{\lower3pt\hbox{\hskip1pt$\sim$}}
		\raise1pt\hbox{$<$}}} 
\def\gsim{\mathrel{\rlap{\lower3pt\hbox{\hskip1pt$\sim$}}
		\raise1pt\hbox{$>$}}} 

\begin{document}

\preprint{APS/123-QED}

\title{Corrections to Landau Fermi-liquid  fixed-point  approximation \\ in  nonlinear bosonized theory: 
Application to $g_A^L$ in nuclei}

\author{Long-Qi Shao}
\email{shaolongqi22@mails.ucas.ac.cn}
\affiliation{School of Fundamental Physics and Mathematical Sciences, Hangzhou Institute for Advanced Study, UCAS, Hangzhou, 310024, China}
\affiliation{Institute of Theoretical Physics, Chinese Academy of Sciences, Beijing 100190, China}
\affiliation{University of Chinese Academy of Sciences, Beijing 100049, China}

	
\author{Mannque Rho}
\email{mannque.rho@ipht.fr}
\affiliation{Universit\'e Paris-Saclay, CNRS, CEA, Institut de Physique Th\'eorique, 91191, Gif-sur-Yvette, France }

\vskip 0.5cm

\date{\today}

\begin{abstract}

We calculated in nonlinear bosonized theory $1/\bar{N}$ corrections to the Landau Fermi-liquid fixed-point (FLFP) axial-vector coupling constant in nuclear matter $g_A^L\approx 1$  to which the Landau parameter $F_1^\omega$ predominantly contributes. 
We obtain the correction to  $F_1^\omega$ to  calculate the correction $\delta g_A^L$ to the axial-vector coupling constant $g_A^L$ at the nuclear saturation density. It comes out to be extremely small, $\delta g_A^L\sim O(10^{-4})$. We discuss how the ``dilaton-limit fixed-point (DLFP)" result $g_A=1$ can be preserved from finite nuclei to high densities relevant to massive neutron stars and its possible impact on $0\nu\beta\beta$ decay processes involved in going beyond the Standard Model.
\end{abstract}

\maketitle

\section{Introduction}


It has been argued~\cite{MRAIQ} that when many-nucleon system is treated as interacting fermions in renormalization-group approach on the Fermi surface, the superallowed Gamow-Teller transition (with momentum transfer $q\approx 0$) is described by the Fermi-liquid fixed-point coupling constant $g_A^L$ captured {\it entirely} by the strong nuclear correlation effects of the quasiparticle on the surface. 
This meant that the long-standing puzzle of the  ``quenched" $g_A^\ast\approx 1$ observed in light nuclei~\cite{Wilkinson:1973zz,gA-review} can be accounted for in terms of a quasiparticle effective axial-vector coupling constant in nuclear effective field theory (EFT) defined by the chiral cut-off scale $\sim 4\pi f_\pi\sim 1$ GeV
\be
g_A^\ast=g_A^L\approx 1.\label{gAL}
\ee
This result should apply to not just light nuclei but also to heavy nuclei and perhaps all the way to massive compact star matter. How this result could impact on nuclear dynamics in general and the search for going beyond the Standard Model will be commented on in the last section.

Up to date there is no {\it ab initio} microscopic many-body calculation to confirm or infirm the prediction (\ref{gAL}).  Highly powerful Monte Carlo calculations~\cite{King:2020wmp} that have been performed for light nuclei with mass number $A\leq 21$ indicate no hint for such a renormalized coupling constant $g_A^\ast$ in many nucleon systems. Modulo possibly small corrections  from $n$-body (for $n\geq 2$) current operators, what could be referred to as {\it fundamental} $g_A=1.276$ can fully account for the axial transitions in nuclear medium. But there is no such fully trustful microscopic calculation available up to date for heavier nuclei.
 
The aim of this paper is to show that the prediction (\ref{gAL}) remains to hold unaffected at least up to $\lsim  n_0$ (where $n_0\approx 0.16$ fm$^{-3}$ is the normal nuclear matter density) under corrections to the Fermi-liquid fixed-point approximation that gave (\ref{gAL}). The strategy we use is to apply to nuclear interactions the technique of nonlinear bosonization of Fermi surfaces anchored on the method of coadjoint orbits developed in condensed matter physics~\cite{Delacretaz:2022ocm}. For the strong interactions involved, we need to implement the degrees of freedom associated with QCD, endowed with both intrinsic and hidden symmetries which bring to the problem complexities absent in, and different from, condensed matter systems. 

The calculation involved consists of three elements. 

The first is the notion that the theory for strong interactions, QCD, gives the nucleons, i.e., light-quark baryons,  as the skyrmions -- and equivalently the weakly-interacting ``constituent  quarks" -- in the limit of large number of colors $N_c\to \infty$.
 Embedded in a skyrmion  medium, the effective quasiskyrmion  mass that we could refer to as ``chiral quasiparticle" mass scales in density as~\cite{br91}\footnote{In this reference, $f^\ast_\chi/f_\chi\approx f_\pi^\ast/f_\pi$ and the latter is given experimentally in deeply bound pionic atom. This relation most likely does not hold in other cases, e.g.,  for dilatonic Higgs models for large $N_f$~\cite{Appelquist:2022mjb}.}
 \be
 m_N^{{\rm skyrmion}\ast}/m_N\approx (g_A^\ast/g_A)^{1/2} f_\pi^\ast/f_\pi.\label{skyrmion-GT}
 \ee 
 Here and in what follows, $\ast$ stands for density dependence. What's taken into account in (\ref{skyrmion-GT}) is, apart from the large $N_c$ limit, scale-invariance of the axial coupling to skyrmions at the classical level and the in-medium dilaton $\sigma_d$ condensate $\la\sigma_d\ra\propto \la\chi\ra$ where $\chi$ here is the linearly scale-transforming ``conformal compensator" field $\chi=f_\chi e^{\sigma_d/f_\chi}$  figuring in the effective field theory Lagrangian. In (\ref{skyrmion-GT}) $f_\pi^\ast/f_\pi$ replaces $f_\chi^\ast/f_\chi$ as explained later. One can identify (\ref{skyrmion-GT}) as an in-medium Goldberger-Treiman relation. It is known that in vacuum, the Goldberger-Treiman relation holds fairly well and we assume that it will also do well in medium.
 
 The next thing we want to establish is the connection between chiral effective field theory for nuclear matter and effective field theory for Landau Fermi-liquid of nucleons.  Extending effective field theory on the Fermi surfaces developed for condensed matter systems~\cite{Shankar:1993pf,Polchinski:1992ed} to nuclear chiral effective field theory incorporating hidden (both vector and scalar) symmetries for strongly interacting nuclear systems~\cite{Friman:1996qc},  it has been shown that
\be
m_L^\ast/m_N\approx \big(1-\tilde{F}_1/3\big)^{-1}\label{landau}
\ee 
where $m_L^\ast$ is the Landau (fixed-point) quasiparticle (a.k.a. nucleon) mass and
\be
\tilde{F}_1=(m_N/m_L^\ast)F_1
\ee 
is the Landau quasiparticle interactions
\be
F_1=F_1^\omega + F_1^\pi,\label{F1}
\ee
where $F_1^{\omega(\pi)}$ stands for the contribution from the Landau quasiparticle interaction in the $\omega(\pi)$  channel. What makes the nuclear Fermi-liquid system strikingly different from the electron Fermi-liquid is the Landau quasiparticle interaction in the pion channel $F_1^\pi$ which will turn out as we will elaborate below to play a crucial role in nuclear properties.

Third, the basic premise of our theory is that the chiral mass (\ref{skyrmion-GT}) valid at the large $N_c$ limit can be equated to the Landau mass (\ref{landau}) in the Fermi-liquid fixed point approximation
\be
 m_N^{{\rm skyrmion}\ast}/m_N\approx m_L^\ast/m_N. \label{mN}
\ee
It follows from (\ref{skyrmion-GT}) that
\be
g_A^\ast\approx g_A \big(1-\frac 13 \Phi^\ast_\chi \tilde{F}_1^\pi)^{-2} \label{FR-gA}
\ee
where
\be
\Phi^\ast_\chi=f_\chi^\ast/f_\chi\approx f_\pi^\ast/f_\pi
\ee
with the last approximate equality assumed to hold in the ``genuine dilaton (GD)" model~\cite{Crewther:2020tgd,Crewther:2015dpa}.

It will be our assertion that one can take the quasi-skyrmion $g_A^\ast$ to be equivalent to the Landau fixed-point quantity $g_A^L$
\be
g_A^L\approx g_A \big(1-\frac 13 \Phi^\ast_\chi \tilde{F}_1^\pi)^{-2}.\label{gAFP}
\ee
It turns out that the product $\frac 13 \Phi^\ast_\chi \tilde{F}_1^\pi$ is more or less independent of density near $n_0$, so one arrives at
\be
g_A^L \approx 1\label{gA1}
\ee
that applies not only to light nuclei but also to nuclear matter at a density $\sim n_0$. What will be surprising is that  $g_A^L\to 1$ exactly at what is called ``dilaton-limit fixed-point (DLFP)"~\cite{Beane:1994ds} at some high density near chiral restoration. 

What takes place in between is the main topic of this paper.  

It will be found that the higher-order corrections found in the framework adopted in this paper to Eq.~(\ref{gA1}) are extremely small, $\sim O(10^{-4})$, even at the normal nuclear matter density $n_0$. What this result, if correct, implies both in the superallowed Gamow-Teller transitions in nuclei as well as in the $0\nu\beta\beta$ transitions relevant for going beyond the Standard Model will be discussed in the Conclusion section.

\section{\label{sec:2}Generalized chiral effective field theory: G$n$EFT}
Our approach is anchored on an effective Lagrangian that incorporates hidden local symmetry (HLS)~\cite{Bando:1984ej,Harada:2003jx} and hidden scale symmetry (HSS)~\cite{Crewther:2020tgd} into chiral EFT mapped to Fermi liquid applicable to nuclear matter (in place of electron systems).  Hidden local symmetry comprising  the lightest vector mesons $\rho$ and $\omega$ is gauge-equivalent to non-linear sigma model~\cite{Bando:1984ej} so can be simply implemented at the classical level (that is at the leading order chiral power counting, $O(p^2)$, in the mesonic sector and $O(p)$ when baryons are coupled).  Hidden scale symmetry is implemented using the conformal compensator field $\chi$ with a suitable dilaton potential $V_d$ appropriate for the GD scale symmetry. For  the problem we are concerned with we do not need to enter the detailed  structure given in the reviews~\cite{Rho:2021zwm,Ma:2019ery}. The formulation made in these reviews involves the mechanism for hadron-quark continuity at a density above $n_0$, typically $\sim 3n_0$, to access the density relevant to the interior of massive compact stars. Although the precise way the cross-over from hadrons to quarks/gluons takes place may not matter quantitatively in  the properties of compact stars, for the problem concerned here at near nuclear matter density, what turns out to be most relevant is the intricate interplay between the coupling of the pions, the $\omega$ and $\sigma_d$ to nucleons. This aspect of the problem has not yet been treated in the literature.

\subsection{\label{sec2} The interplay between $\omega$ meson coupling and $\sigma_d$ meson coupling}
In arriving at Eq.~(\ref{FR-gA}), what is involved is an interplay between the $\omega$ and $\sigma_d$ mesons in nuclear matter treated as a Fermi liquid  on the Fermi surface. For this matter, what's most important is the notion of the ``genuine dilaton" (GD for short) in QCD as put forward by Crewther and collaborators~\cite{Crewther:2020tgd,Crewther:2015dpa}. The GD scheme is characterized by the assumption that there exists  an infrared fixed point (IRFP) $\alpha_{IR}$ for $N_f\leq 3$ at which both scale symmetry and chiral symmetry (in the chiral limit) are realized in the Nambu-Goldstone (NG) mode, populated by the massless NG bosons $\pi$ and dilaton $\sigma_d$ whose decay constants are non-zero. The characteristic of this notion that we espouse is that it accommodates the massive nucleons $\Psi$ and vector mesons $V_\mu=(\rho, \omega)$ at the IR fixed point. 

The GD in a chiral Lagrangian  plays not only the role of reducing the nucleon mass from $m_N$ to $m_N^{s}$ by the attractive coupling to the nucleon like $\sigma$ meson in Walecka mean field theory~\cite{Walecka:1974qa} but also endows the BR scaling~\cite{Brown:1991kk} caused by the vacuum change by nuclear medium
\begin{equation}
	\frac{m_V^\ast}{m_V}\approx\frac{m^\ast_s}{m_s}\approx\frac{f_\pi^\ast}{f_\pi}\equiv \Phi_\chi^\ast, \label{eq:BR scaling relation}
\end{equation}
where $V$ means vector meson, $m_V$ the vector meson mass, $m_s$ the scalar (either Walecka's $\sigma$ or $\sigma_d$ in G$n$EFT) meson mass. 

In mapping to Fermi liquid, we integrate out the $\omega$ and $\pi$ mesons, leaving local four-fermion and non-local four-fermion interactions, respectively ($\rho$ channel will not figure in the calculation of $g_A^L$, but will figure in the calculation of the gyromagnetic ratio mentioned below). The quasiparticle energy is 
\begin{equation}
	\varepsilon(p)=\frac{p^2}{2m_N^{\sigma}}+C_\omega^2 n+\Sigma_\pi(p), \label{epsilon}
\end{equation}
where $m_N^{\sigma}$ is the BR-scaled nucleon mass and  $C_\omega^2=g_\omega^2/m_\omega^{\ast2}$ with $g_\omega$ the $\omega$-N coupling constant. $\Sigma_\pi(p)$ is self-energy given by the pionic Fock term. Taking derivatives with respect to momentum at the Fermi surface
\begin{equation}
\frac{d\varepsilon(p)}{dp}\bigg|_{p=p_F}=\frac{p_F}{m_L^\ast}=\frac{p_F}{m_N^\sigma}+\frac{d\Sigma_\pi(p)}{dp}\bigg|_{p=p_F},
\end{equation}
where $p_F$ is the Fermi momentum. Eq.~(\ref{skyrmion-GT}), Eq.~(\ref{landau}), Eq.~(\ref{mN}), and the Fock term at $n_0$~\cite{Brown:1980bg}
\begin{equation}
	\tilde{F}_1^\pi=-3\frac{m_N}{p_F}\frac{d\Sigma_\pi(p)}{dp}\bigg|_{p=p_F}  
\end{equation}
lead us to identify $g_A^L\approx 1$, Eq.~(\ref{gAFP}),  as a Landau fixed point quantity. A quantity that figures importantly to justify this is the relation
\begin{align}
	\frac{m_N}{m_N^\sigma}&=\Phi_\chi^{\ast-1}=1-\frac{1}{3}\tilde{F}_1^\omega \label{omega sigma}.
\end{align}
{
The quantities involved here are taken to be Landau-fixed point (LFP) quantities.  What we will calculate below as $1/\bar{N}$ corrections deal with the LFP quantities and the relation  (\ref{omega sigma}) correlates the attractive interaction effect associated with the dilaton, i.e., the BR scaling, and the repulsion due to $\omega$ exchange. Although we have not fully understood -- a task we have left for the future work, we think it is this relation that dictates intricately the size of the $1/\bar{N}$ corrections estimated in this paper.

\subsection{Mesonic fields}
As stated, in formulating renormalization group approaches to interacting nucleons, it is more astute to introduce mesonic fields {\it ab initio} that are massive such as the light-quark vector mesons $(\rho,\omega)$ in addition to the pseudo-Nambu-Goldstone bosons $\pi$, $\sigma_d$ etc. For very low energy excitations or densities, the bosonic fields can be integrated out with higher derivative terms appearing. In fact even the pions can be integrated out leading to pionless EFT which can be treated in terms of the power series. The pionless EFT will breakdown if the excitation involves the scale higher than the pion mass. The standard chiral EFT (S$\chi$EFT) anchored on the chiral Lagrangian with nucleons and pions has been shown to work well up to the scale corresponding to the scale of the masses of the mesons integrated out, say $O(m_\rho)$.  The power expansion goes typically to N$^{q}$LO for $q\leq 3$ and is to breakdown when the scale involved goes above $m_\rho$. The S$\chi$EFT therefore is expected not to work for compact-star densities where the density involved is $\sim (4-7)n_0$.}

The approach mentioned in the previous subsection that maps G$n$EFT Lagrangian to the Landau-Fermi-liquid fixed point theory of many-nucleon systems can circumvent such difficulty. When the nucleons are put on the Fermi surface, the Fermi-liquid fixed point (FLFP) approximation corresponds to taking $1/\bar{N}$ to zero, where $\bar{N}=p_F/\Lambda$ with $\Lambda$ the cut-off with respect to the Fermi surface. It becomes more reliable as density increases. When we consider $\Lambda$ to be a finite quantity small compared to $p_F$, the $1/\bar{N}$ correction should enter into the fixed point result Eq.~(\ref{gAL}). This is the double-decimation procedure applied in $V_{\rm lowK}$ RG in finite nuclei~\cite{vlowk}. The main result of this paper is that Eq.~(\ref{gAL}) still holds when $1/\bar{N}$ correction enters.

Before going into the $1/\bar{N}$ corrections, it should be noted that the pion and $\omega$ contribute differently to the FLFP result. To illustrate this, we explain how the gyromagnetic ratio for the proton in heavy nuclei $g^p_l$ comes out in the FLFP approximation in G$n$EFT.  The calculation is essentially the same as the simplified chiral Lagrangian with HLS fields implemented with the hidden scale symmetry used in \cite{Song:2000cu}.  We will therefore simplify the discussion, leaving the details to \cite{Song:2000cu}.

In G$n$EFT, the Lagrangian taken at the mean-field (viz, FLFP) approximation~\cite{Shankar:1993pf,Friman:1996qc}  gives  the Migdal formula~\cite{Migdal} for the quasiparticle convection current 
\be
\boldsymbol{J}=\frac{\boldsymbol{p}}{m_N}g_l=\frac{\boldsymbol{p}}{m_N}\Big( \frac{1+\tau_3}{2}+\frac{1}{6}(\tilde{F}_1-\tilde{F}_1^{'} )\tau_3\Big),
\ee
where $\tilde{F}_1=\tilde{F}_1^\pi+\tilde{F}_1^\omega$, $\tilde{F}_1^{'}=\tilde{F}_1^{'\pi}+\tilde{F}_1^{'\rho}$, and $g_l=(1+\tau_3)/2+\delta g_l$, where $\delta g_l$ is the anomalous gyromagnetic ratio $\delta g_l=\delta g_l^0 +\delta g_l^1$ with $(0,1)$ standing for (isoscalar, isovector). It was found~\cite{Friman:1996qc}  that
\be 
\delta g_l^0 &=&0,\label{isoscalar} \\
\delta g_l^1 &=&  \frac{4}{9}\large[\Phi_\chi^{\ast-1}-1-\frac{1}{2}\tilde{F}_1^\pi]\tau_3.\label{isovector}
\ee
It turns out that the prediction (\ref{isovector}) agrees precisely with the available experiment~\cite{Nolte:1986ghy}.

Now going from the EFT chiral Lagrangian to the Migdal formula is highly nontrivial, the details of which are found in \cite{Friman:1996qc,Song:2000cu}. The intricacy comes in two ways. First,  the nucleon mass that figures in the Lagrangian is the BR scaled mass $m_N^\sigma$, so the single-nucleon convection current will be $ {\boldsymbol{J}}_{1-body}=\frac{\boldsymbol{p}}{m_N^\sigma}\frac{1+\tau_3}{2}$. As explained in \cite{Song:2000cu}, this would violate the $U_{EM} (1)$ gauge invariance. Second, the same holds when $m_N^\sigma$ is replaced by the Landau mass $m_L^\ast$. This problem known in condensed matter systems as ``Kohn effect" can be remedied when two-body exchange currents involving the vector meson exchange currents are taken into account~\cite{Song:2000cu}. What figures there is what's known as ``back-flow current" that is required by Ward identity.  It effectively restores the $U(1)$ gauge invariance. This means that whatever figures as corrections in $F_1^\omega$ should be constrained by the gauge invariance involving Ward identity. 

{Now the  $U(1)$ gauge invariance says nothing about the isovector part, Eq.~(\ref{isovector}), so there is no such constraint in the $\rho$ channel. However, in deriving Eq.~(\ref{isovector}), the nonet relation $C_\rho^2=C_\omega^2/9$ was used, so the $\rho$ channel contributes in the same way as $\omega$ channel except for the coupling constant and the isospin matrix; so we have $\tilde{F}_1^{'\rho}=\tilde{F}_1^{\omega}/9$, which also holds when $1/\bar{N}$ corrections are included. The contribution of the $\rho$ channel in Eq.~(\ref{isovector}) was represented by $\Phi_\chi^\ast$ through Eq.~(\ref{omega sigma}). The fact that Eq.~(\ref{isovector}) agrees well with experiments tells us that the corrections to $\tilde{F}_1^\pi$ and $\tilde{F}_1^{'\rho}$ should cancel each other significantly or both be small, otherwise the good result Eq.~(\ref{isovector}) will be spoiled. The above theory based on the chiral Lagrangian implemented with hidden local and scale symmetries in the FLFP approximation gives satisfying results for $g_A^L$ and $\delta g_l^p$. The renormalizations of the vertices involved are in one-to-one correspondence with the nucleon self-energy except for the isospin index. Given that the $\tilde{F}_1^{'\rho}$ is closely related to ${F}_1^{\omega}$, whose correction turn out to be small, we will assume the possible corrections to $F_1^\pi$ are ignorable, i.e. corrections to $\tilde{F}_1^\pi$ and $\tilde{F}_1^{'\rho}$ are both small to preserve the good result for gyromagnetic ratio.

What is important in our problem concerned here is the $\omega$ meson contribution to the BR-scaled nucleon mass (\ref{omega sigma}). At the mean-field level (that is, in the FLFP approximation), it is connected to the universal BR scaling $\Phi_\chi^\ast$. However, if one goes beyond the FLFP approximation -- that is higher order in $1/\bar{N}$ -- $\tilde{F}_1^\omega$ will be modified and as a consequence, the nucleon mass will scale differently from the universal BR-scaling $\Phi_\chi^\ast=f_\pi^\ast/f_\pi$. 

We denote the modification as the nucleon mass shift associated with the $\omega$ exchange 
\be
m_N^\sigma\to m_N^\sigma +\delta m_N^\omega.\label{deltamN}
\ee
It is important to note that this modified scaling should not affect the $U(1)$ gauge invariance. 

\subsection{$1/\bar{N}$ corrections}

Now let us return to the $g_A$ problem (\ref{gAFP}). We would like to see  how $g_A^L\approx 1$ is modified by corrections to the FLFP approximation and also by the increase of  density beyond the normal matter density $n_0$.  Let us mention that the $g_A^L$ at higher density is not an academic curiosity. In fact, as already mentioned, it is closely linked to the possible property of the GD at what is termed DLFP at the density at which $\la\chi\ra$ goes to zero~\cite{Beane:1994ds}. It is also connected to the possible precocious onset of the pseudo-conformal sound speed~\cite{Ma:2019ery}. What's at issue in the behavior of $g_A^L$ is the product of the two quantities $\tilde{F}_1^\pi$ and $\Phi_\chi^\ast$, which also figure in the convection current discussed above. The question then is how these quantities are modified by the corrections. 

We first consider $\tilde{F}_1^\omega$.  From Eq.~(\ref{omega sigma})
\begin{align}
	\frac{m_N}{m_N^\sigma}=1-\frac{1}{3}\tilde{F}_1^\omega
\end{align}
We see that the scaling of the nucleon mass by the BR scaling capturing the role of the attraction brought by the dilaton field is tied to the property of the $\omega$ contribution to the Landau parameter. At the Fermi-liquid fixed point, it is given by the universal BR scaling factor $\Phi_\chi$. Going beyond the FLFP approximation, it will depend on the vector meson ($\rho$ or $\omega$) exchanged. Here it will be the $\omega$ channel whereas in $\delta g_l^1$,  Eq.~(\ref{isovector}), it will be the $\rho$ channel. In principle they could be different.

Now we focus on calculating $\delta m_N^\omega$ in Eq.~(\ref{deltamN}). For this we need the standard textbook notations for Landau Fermi-liquid theory applied to nuclear systems. The variation of the energy of the system is given by
\begin{equation}
    \delta E=\sum_{\boldsymbol{p}}\varepsilon_{p}\delta f(\boldsymbol{p}) +\frac{1}{2V}\sum_{\boldsymbol{p},\boldsymbol{p'}}\mathcal{F}(\boldsymbol{p},\boldsymbol{p'})\delta f(\boldsymbol{p})\delta f(\boldsymbol{p'}),
\end{equation}
where $\varepsilon_p$ is the quasi-particle energy, $\delta f(\boldsymbol{p})$ is the deviation of the fermion occupation number from the ground state, $V$ is the volume, $\mathcal{F}(\boldsymbol{p},\boldsymbol{p'})$ are quasi-particle interactions on the Fermi surface,
\begin{equation}
  \begin{aligned}
     \mathcal{F}(\boldsymbol{p},\boldsymbol{p'})=&f(\cos\theta)+f'(\cos\theta)(\boldsymbol{\tau}\cdot\boldsymbol{\tau'})+g(\cos\theta)(\boldsymbol{\sigma}\cdot\boldsymbol{\sigma'}) \\&+g'(\cos\theta)(\boldsymbol{\tau}\cdot\boldsymbol{\tau'})(\boldsymbol{\sigma}\cdot\boldsymbol{\sigma'})
  \end{aligned}
\end{equation}
with
\begin{equation}
    f(\cos\theta)=\sum_{l=0}^\infty f_{l} P_l (\cos\theta), \quad f'(\cos\theta)=\sum_{l=0}^\infty f'_{l} P_l (\cos\theta),
\end{equation}
where $P_l (\cos\theta)$ are Legendre polynomials. The dimensionless Landau parameters we use are 
\begin{equation}
    F_l=Nf_l, \quad F'_l=Nf'_l
\end{equation}
where $N$ is the fermion density at the Fermi surface
\begin{equation}
    N=\frac{\nu m_L^\ast p_F}{2\pi^2}, \label{eq:fermion density}
\end{equation}
where $\nu$ is degeneracy factor, and $m_L^\ast$ the Landau effective mass. 

\subsubsection{Fermi-liquid fixed-point approximation}

We first calculate $F_1^\omega$ in the Landau Fermi-liquid fixed point approximation.  This can be done by first reducing the G$n$EFT Lagrangian to Walecka's linear $\sigma/\omega$ model~\cite{Walecka:1974qa,Matsui:1981ag} 
\begin{equation}
 \begin{aligned}
\mathcal{L}=&\bar{\Psi}(i\slashed{\partial}-m_N)\Psi+\frac{1}{2}(\partial_\mu\sigma\partial^\mu\sigma-m_s^2\sigma^2)\\ &-\frac{1}{2}(\frac{1}{2}\omega_{\mu\nu}\omega^{\mu\nu}-m_\omega^{\ast2}\omega_\mu \omega^\mu)+g_s\bar{\Psi}\Psi\sigma-g_\omega\bar{\Psi}\gamma_\mu\Psi \omega^\mu
 \end{aligned}
\label{fermion lagrangian}
\end{equation}
and then doing relativistic mean-field calculation. It corresponds to taking one-loop term in Fig.~\ref{fig:iterated 1-loop}
\begin{figure}[!ht]
    \centering
    \includegraphics[width=0.5\textwidth]{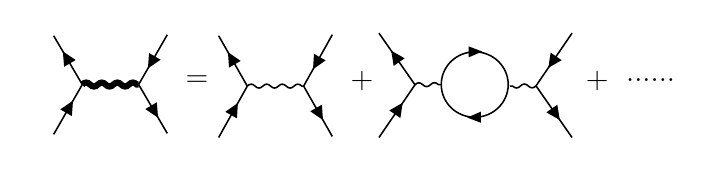}
    \caption{\label{fig:iterated 1-loop}Two-body interactions by $\omega$ exchange;  the thick wavy line represents the full $\omega$ propagator, the thin wavy line the free $\omega$ meson propagator, the solid line the nucleon with Walecka effective mass $m_N^\sigma$. }
\end{figure}

From Fig.~\ref{fig:iterated 1-loop} one can write down the two-body nuclear interaction as

\begin{widetext}
  \begin{equation}
f^\omega(\boldsymbol{p},\boldsymbol{p'})=\bar{u}(p)\gamma^\mu u(p)D_\omega\bar{u}(p')\gamma_\mu u(p')+\bar{u}(p)\gamma^\mu u(p)D_\omega\Pi_{\mu\nu}D_\omega\bar{u}(p')\gamma^\nu u(p')+\cdots, \label{eq:2-body interaction}
  \end{equation}
\end{widetext}
where $D_\omega$ is the free $\omega$ meson propagator 
\begin{equation}
    D_\omega=\frac{g_\omega^2}{m_\omega^{\ast2}-q_\mu q^\mu}. \label{eq:omega propagator}
\end{equation}
and $\Pi_{\mu\nu}$ is the polarization tensor of $\omega$ meson field.
Taking in the polarization tensor $\Pi_{\mu\nu}$  the forward scattering limit $|\boldsymbol{q}|/q_0\rightarrow0$ one finds that only the $\Pi_{ii}$ component is non-vanishing
 \begin{equation}
    \Pi_{ii}\approx \frac{n}{E_F}\label{eq:fermionic polarization tensor}
\end{equation}
where $E_F=\sqrt{m_N^{\sigma2}+p_F^2}$ and $n$ is the nuclear matter density. We equated the shifted scalar mass in the mean-field calculation to the BR-scaled nucleon mass. Using Eq.~(\ref{eq:omega propagator}), Eq.~(\ref{eq:fermionic polarization tensor}), and the spinor $u$,
 we get $F_1^\omega$ (with the Legendre polynomial $P_1(\cos \theta)=\cos \theta$),
\begin{equation}
    \begin{aligned}
        \frac{1}{3}F_1^\omega&=\frac{1}{3}f_1\cdot N \\&=-\frac{1}{3}C_\omega^2\frac{p_F^2}{E_F^2(1+C_\omega^2\Pi_{ii})}\cdot\frac{\nu p_F m_L^\ast}{2\pi^2}.  \label{eq:F_1omega}
    \end{aligned}   
\end{equation}
We see that it is the polarization tensor $\Pi_{ii}$ that controls  $F_1^\omega$. To reproduce the FLFP result in~\cite{Friman:1996qc}, we resort to  the non-relativistic form
\begin{equation}
    \Pi_{ii}=\frac{n}{m_N^\sigma},\quad \frac{1}{3}F_1^\omega=-C_\omega^2\frac{nm_L^\ast}{m_N^\sigma(m_N^\sigma+C_\omega^2\Pi_{ii}m_N^\sigma)}  \label{eq:polarization tensor in NR}
\end{equation}

\subsubsection{\label{sec:level2.2}To go beyond the FLFP approximation}
It is possible to go beyond the FLFP approximation by using the $V_{lowk}$ renormalization-group approach~\cite{vlowk}. We find it however easier and more systematic to apply the nonlinear bosonized theory (NBT)~\cite{Delacretaz:2022ocm}. We will first reproduce the polarization tensor in Eq.~(\ref{eq:polarization tensor in NR}) using the NBT and then calculate the corrections to the  polarization tensor denoted  $\delta \Pi_{ii}$.

The nonlinear bosonized theory uses bilinear fermion operators as exponential of bosonic degrees of freedom which can be seen as particle-hole excitations on the Fermi surface. The Lagrangian  is 
\begin{equation}
    S=\int dt\left\langle f_0,U^{-1}(\partial_t-\varepsilon)U \right\rangle  \label{eq:NB lagrangian}
\end{equation}
where $f_0(\boldsymbol{p})=\Theta(|\boldsymbol{p}|-p_F)$ is the ground state fermion occupation number and $U=\exp(-\phi)$ with $\phi$ the dynamical bosonic degree of freedom that characterizes nonlinear structure of Fermi surface, $\phi$ lives on both momentum space and coordinate space. 
\begin{equation}
    \int dt\left\langle A,B \right\rangle=\int dt d\boldsymbol{x}d\boldsymbol{p} A\cdot B,
\end{equation}
\begin{equation}
    U^{-1}AU=A+\{\phi,A\}+\frac{1}{2!}\{\phi,\{\phi,A\}\}+\cdots,
\end{equation}
where $\{,\}$ is the Poisson bracket.

The first term in Eq.~(\ref{eq:NB lagrangian}) is the Wess-Zumino-Witten (WZW) term that captures the geometric phase when the Fermi surface evolves with time, while the second term is the normal quasiparticle energy term. For convenience, we use the non-relativistic $\varepsilon=p^2/2m_N$.

As in Sec.~\ref{sec:2},   we incorporate $\sigma$ and $\omega$ fields, use Walecka's mean-field approximation by integrating out the $\sigma$ meson and  change $m_N\rightarrow m_N^\sigma$ (which can be considered as the BR-scaled mass)
\begin{equation}
        S=\int dt\left\langle f_0,U^{-1}(\partial_t-\omega_0-\frac{(\boldsymbol{p}+\boldsymbol{\omega})^2}{2m_N^\sigma})U \right\rangle+S_\omega  \label{eq:bosonic lagrangian}
\end{equation} 
where $S_\omega$ is the $\omega$ meson sector, $\omega_0$ and $\boldsymbol{\omega}$ are the four component $\omega_\mu$. 

With $U$ expanded in different orders in $\phi$, we can calculate the $\omega$ propagators systematically. The details of the power counting and calculations are  given in Appendix. The leading-order corrections to the free $\omega$ propagators are shown in Fig.~\ref{fig:omega propagator}.
\begin{figure}[hb]
    \centering
    \includegraphics[width=0.5\textwidth]{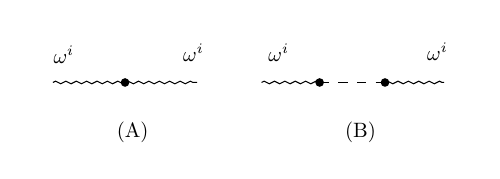}
    \caption{\label{fig:omega propagator}The leading-order corrections to the free $\omega$ propagator, the dashed line represents the bosonic field $\phi$, the wavy line the spatial component of the $\omega$ meson. The black dots represent different vertices in the Lagrangian.}
\end{figure}

As in the fermion description, we are interested in $\Pi_{ii}$, so only the spatial components of the $\omega$ meson are involved. The leading-order corrections, the details of which are relegated to Appendix B, are
\begin{equation}
    \Pi_{ii}=-\frac{\nu p_F^3}{6\pi^2m_N^\sigma}\frac{m_N^\sigma|q_0|}{2p_F|\boldsymbol{q}|}\log\frac{m_N^\sigma|q_0|-p_F|\boldsymbol{q}|}{m_N^\sigma|q_0|+p_F|\boldsymbol{q}|}.  \label{eq:polarization tensor}
\end{equation}
In the limit $|\boldsymbol{q}|/q_0\rightarrow0$, $\Pi_{ii}=n/m_N^\sigma$, the same as Eq.~(\ref{eq:polarization tensor in NR}).

The next-order diagrams are $1/\bar{N}=\Lambda/p_F$ suppressed. There are more diagrams at this order.  However it turns out that in the limit $q_0\rightarrow0$, $|\boldsymbol{q}|\rightarrow0$ and $|\boldsymbol{q}|/q_0\rightarrow0$, only one diagram Fig.~\ref{fig:next leading order correction} survives.
\begin{figure}[h]
    \centering
    \includegraphics[width=0.3\textwidth]{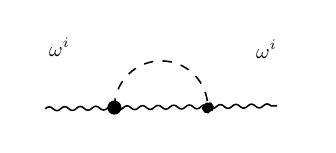}
    \caption{\label{fig:next leading order correction}Next-order corrections to the free $\omega$ meson propagator}
\end{figure}

Notice that the nonlinear structure has not yet entered in the diagrams we have dealt with so far. The reason is that no vertices that connect two dashed lines appear in these diagrams. Such vertices will appear at  higher orders or if we loosen  the constraint that $q_0$ and $|\boldsymbol{q}|$ approach zero in the way $|\boldsymbol{q}|/q_0\rightarrow0$ as shown in Fig.~\ref{fig:higher order}.
\begin{figure}[hb]
    \centering
    \includegraphics[width=0.5\textwidth]{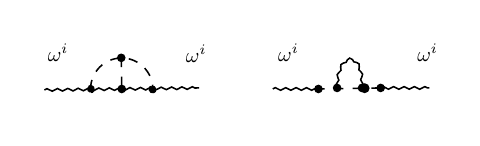}
    \caption{The left graph is the higher-order correction to the $\omega$ meson propagator, the right graph is the next-to-leading order correction.}
    \label{fig:higher order}
\end{figure}

The calculation of Fig.~\ref{fig:next leading order correction} involves large momentum transfer~\cite{Shankar:1993pf} as presented in Appendix \ref{app:2.2}. We will designate this diagram as $1/\bar{N}$ correction to the $\omega$ meson propagator and also to $F_1^\omega$. The result is (see Appendix B)
\begin{equation}
    \delta \Pi_{ii}=-\frac{\nu\Lambda g_\omega^2}{64\pi^4m_N^{\sigma2}}\int_\Lambda^{2p_F} dk\frac{k^4}{k^2+m_\omega^{\ast2}-\frac{k^2\sqrt{k^2+m_\omega^{\ast2}}}{2m^\sigma_N}}.  \label{eq:delta Pi}
\end{equation}
We have assumed that the $\omega$-meson-exchange Landau parameter $F_1^\omega$ and the pion-exchange $F_1^\pi$ are not correlated at the order considered as at the FLFP limit~\cite{Friman:1996qc}. Since we are at the fixed point at $n$ near $n_0$, we consider this to be reasonable.

\subsubsection{$1/\bar{N}$ correction to   $g_A^L$}

We have derived the $1/\bar{N}$ corrections to the $\omega$ propagators in Sec.~\ref{sec:level2.2}. The corrections to the $\omega$ meson propagators will directly modify the $F_1^\omega$ induced by the exchange of the $\omega$ meson. According to the argument made in Sec.~\ref{sec2}, Eq.~(\ref{omega sigma}) is essential for preserving gauge invariance. We propose as in Eq.~(\ref{deltamN}) that since Eq.~(\ref{epsilon}) corresponds to the FLFP result, when we are away from the FLFP, $m_N^\sigma$ should be replaced by $m_N^\sigma+\delta m_N^\omega$. We define
\begin{equation}
    \zeta=\frac{m_N^\sigma+\delta m_N^\omega}{m_N}.
\end{equation}
The quasiparticle energy  becomes
\begin{equation}
    \varepsilon(p)=\frac{p^2}{2(m_N^\sigma+\delta m_N^\omega)}+C_\omega^2 n+\Sigma_\pi(p).  \label{eq:quasi-particle energy2}
\end{equation}
Therefore in Eq.~(\ref{gAFP}) and Eq.~(\ref{omega sigma}), we need to replace  $\Phi_\chi^\ast$ by $\zeta$
\begin{align}
    \frac{m_N}{m_N^\sigma+\delta m_N^\omega}&=\zeta^{-1}=1-\frac{1}{3}\tilde{F}_1^\omega, \label{eq:omega sigma relation2} \\ 
    \frac{g_A^L+\delta g_A^L}{g_A}&=(1-\frac{1}{3}\zeta \tilde{F}_1^\pi)^{-2}. \label{eq:g_a quenching2}
\end{align}
where $\delta g_A^L$ is the $1/\bar{N}$ correction to the axial coupling constant in medium.

Using Eq.~(\ref{eq:polarization tensor in NR}) with the corrections included, we have
\begin{equation}
    \zeta^{-1}=1+\frac{C_\omega^2 n m_N}{m_N^{\sigma}(m_N^\sigma+C_\omega^2(\Pi_{ii}+\delta \Pi_{ii})m_N^\sigma)}. \label{eq:omega sigma relation3}
\end{equation}

Finally ignoring higher-order corrections to $F_1^\pi$ that we argue to be ignorable, we can make an estimate of the $1/\bar{N}$ correction to $g_A^L$.

With the degeneracy factor $\nu=2$ and $g_\omega=10.15$ from Eq.~(\ref{omega sigma}), taking the cutoff on top of the Fermi surface to be $\Lambda\sim 10 $ MeV for illustration,  we obtain with (\ref{eq:delta Pi}) 
\be
\delta g_A^L=1.8\times10^{-4}\ \ {\rm at} \ n=n_0\simeq 0.16 {\rm fm} ^{-3}.
\ee 
{The correction is extremely small. One can see roughly  how it comes about as follows:{ The denominator in the integral of Eq.~(\ref{eq:delta Pi}) in the range  $m_\omega^{\ast2} -1.3m_\omega^{\ast2}$  giving the integrand  $\sim k^4/m_\omega^{\ast2}m_N^{\sigma2}$ comes out always less than 1 in the integral region. Multiplied by the remaining factor $\Lambda g_\omega^2/32\pi^4$, $\delta\Pi_{ii}$ turns out to be a tenth of MeV$^2$, while $\Pi_{ii}=\nu p_F^3/6\pi^2m_N^\sigma \sim 10^3$ MeV$^2$. The quantity $\zeta$ is insensitive to $\Pi_{ii}$ given that the leading-order $\omega$ exchange contribution without polarization tensor (first diagram on the RHS of Fig.~(\ref{fig:iterated 1-loop})) is the main contribution to the Landau parameter $F_1^\omega$. Reflecting on the difference between $\zeta$ and $\Phi_\chi^\ast$ and $\delta g_A^L$ (Eq.~(\ref{eq:g_a quenching2})), the correction to $g_A^L$ becomes of $\sim O(10^{-4})$.} The dependence of $\delta \Pi_{ii}$ on the cutoff $\Lambda$ is almost linear, because the integrand in Eq.~(\ref{eq:delta Pi}) contribute little when $k$ is small. Therefore  $\delta g_A^L$ is almost linear in $\Lambda$ as well, as seen in Fig.~\ref{fig:delta ga to Lambda}.}
  \begin{figure}[h]
    \centering
    \includegraphics[width=0.45\textwidth]{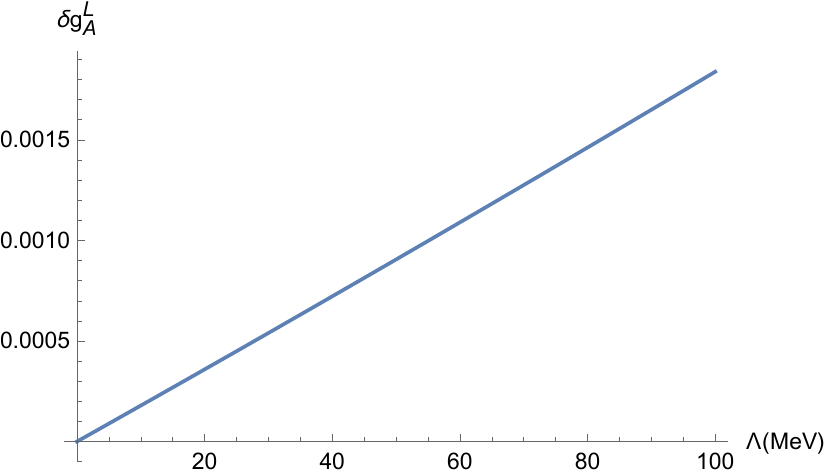}
    \caption{\label{fig:delta ga to Lambda}Dependence of $\delta g_A^L$ on the cutoff $\Lambda$}
\end{figure}

Now if one applies the same reasoning to the anomalous gyromagnetic ratio $\delta g_l^i$, one finds that
 the inclusion of $1/\bar{N}$ correction also receives small corrections $\sim O(10^{-4})$ by substituting $\Phi_\chi^\ast$. This would confirm that the pion contribution should also be unafffected in $\delta g_l^i$.

\section{conclusion}
In this work, we calculated $1/\bar{N}$ corrections to $\omega$-meson propagators using the nonlinear bosonization of Fermi surface approach, and mapped the corrections to the FLFP result $g_A^L$ at $n_0$. We ignored  possible corrections to the pionic Fock term on the ground that its effect in the anomalous gyromagnetic ratio subject to a similar pionic  correction can be safely ignored. We show that the corrections $\delta g_A^L$ are of order $10^{-4}$ and will tend to go to zero reaching the dilaton-limit fixed point. This result is connected with the intricate interplay mentioned above between the attraction due to the scalar excitation, i.e., dilaton, and the repulsion due to the $\omega$ excitation effective from low density to high density. The nonlinear structure of the Fermi surface in nuclear correlations does not enter into the leading-order and next-to-leading-order corrections to the $\omega$ meson propagator. This feature of the nonlinear bosonized theory (shown graphically in Appendix.~\ref{app:2.2}) may be seen as a result of the unscaling of $g_{\omega NN}$ in one-loop order in~\cite{Paeng:2013xya}, which is important for nuclear matter to be stable in the ``half-skyrmion" phase.

It is highly notable that $g_A^L$ remains more or less unaffected by density from finite nuclei to dense baryonic matter. A similar phenomenon seems to be in action in the pseudo-conformal sound velocity $v_s^2/c^2\approx 1/3$ in dense compact-star matter at $n\gsim 3n_0$. As suggested in \cite{MRAIQ}, it could also reflect a hidden scale symmetry in action in nuclear correlations.

That $g_A^L$ in nuclear matter is close to 1 has two important implications in  physics that have not been recognized up to date: One in nuclear dynamics and the other in going beyond the Standard Model. 

As for the first, the question is what  $g_A^L$ represents in what's observed in nuclear weak processes. The $g_A^L$ as phrased in terms of Landau Fermi-liquid theory is the axial coupling constant  $g_A$ with which a quasiparticle on top of the Fermi sea makes the zero momentum  and zero energy transfer Gamow-Teller transition, such that ${q}/\omega \to 0$, taking place in infinite nuclear matter. In real nuclear processes involving finite nuclei as discussed in \cite{Wilkinson:1973zz,gA-review}, phrased in shell model, the closest to the Fermi-liquid result is the ``Extreme Single Particle Shell Model (EPSM)" applicable in superallowed Gamow-Teller transitions in doubly magic-shell nuclei. The heaviest nuclear system  so far studied  is  the $^{100}$Sn nucleus.  It has been argued that the constant $g_A^L$ obtained in the Fermi-liquid model can be closely mapped to what enters in the superallowed Gamow-Teller transition where a proton in the filled proton magic shell  makes the transition to the lowest neutron in the empty neutron magic shell with $q/\omega \to 0$.  Ideally it should  involve a single unique daughter state. In reality the daughter state may not be unique. However with as small an uncertainty as in the doubly magic-shell nucleus, $g_A^L$ should correspond closely to $q$ times  $g_A$ where $g_A$ is the axial coupling constant in (free-space) neutron beta decay and $q$ is the quenching factor~\cite{gA-review}. What this means is that $q$ should capture the {\it full or exact} nuclear correlations leading to the full quenching factor that gives the ``effective axial constant" $g_A^{eff}\approx 1$ observed in nature. In nuclear physics, this implies  that if one were to do the ``full calculation" in the sense of  \cite{Wilkinson:1973zz},  the axial coupling constant applicable in nuclear effective field theory defined at the chiral symmetry scale $\Lambda_{\chi}\approx 4\pi f_\pi\sim 1$ GeV should be the free-space value $g_A=1.276...$ measured in neutron $\beta$ decay. 

Secondly, $g_A^L\approx 1$ with $q\approx 0.78$ can have a big impact on the effort to go beyond the Standard Model by measuring $0\nu\beta\beta$ processes in nuclei.  It appears that there can be a non-trivial renormalization of $g_A$ in heavy nuclei induced by ``quantum anomaly"~\cite{MRAIQ} coming from the degrees of freedom integrated out of the chiral symmetry scale. Up to date, there has been no ``smoking-gun" indication for a ``fundamental" quenching of $g_A$ in {\it any} nuclear processes, but the most recent measurement -- heralded to have been improved -- at RIKEN~\cite{Lubos:2019nik} in the superallowed GT transition in the doubly magic-shell nucleus $^{100}$Sn, if confirmed, would indicate the basic axial-vector coupling constant $g_A$ would undergo a quenching as big as 30-40 \% reduction in {\it all} nuclear weak processes, this  independent of the nuclear correlation effects associated with $g_A^L$. This would imply a huge reduction in the decay rate in $0\nu\beta\beta$ processes for going beyond the SM. This calls for revisiting the $^{100}$Sn beta decay process, both experimentally and theoretically.
  

\appendix

\section{\label{app:1}Power counting of the nonlinear bosonized theory}

The action of the nonlinear bonsonized theory coupled with mesons can be written schematically as 
\begin{equation}
    S=\int_{\boldsymbol{x},\boldsymbol{p}} dt\left\langle f,\partial_t-\omega_0-\frac{(\boldsymbol{p}+\boldsymbol{\omega})^2}{2m_N^\sigma} \right\rangle+S_\omega, \label{eq:bosonized lagrangian2}
\end{equation}
where
\begin{equation}
    \begin{aligned}
        f=&f_0-\{\phi,f_0\}+\frac{1}{2}\{\phi,\{\phi,f_0\}\}-\cdots  \\=&\Theta(p_F-|\boldsymbol{p}|)+\delta(|\boldsymbol{p}|-p_F)\boldsymbol{n}_\theta\cdot\nabla_x\phi+\cdots
    \end{aligned}
\end{equation}
The fermion density is
\be
n(t,\boldsymbol{x})=\nu\int _{\boldsymbol{p}}f=\frac{\nu p_F^3}{6\pi^2}+\frac{\nu p_F^2}{(2\pi)^3}\int d^2\theta \boldsymbol{n}_\theta\cdot\nabla\phi(t,\boldsymbol{x},\theta)+\cdots. \nonumber \\  \label{eq:fermion occupation number}
\ee

In the action expanded in terms of $\phi$, the free propagator of $\phi$ comes out
\begin{equation}
    \left\langle \phi\phi'\right\rangle(q_0,\boldsymbol{q})=i\frac{(2\pi)^3}{p_F^2}\frac{\delta^2(\theta-\theta')}{\boldsymbol{n}_\theta\cdot\boldsymbol{q}(q_0-\frac{p_F}{m_N^\sigma}\boldsymbol{n}_\theta\cdot\boldsymbol{q}+i\epsilon)}. \label{eq:phi propagator}
\end{equation}
The details are relegated to~\cite{Delacretaz:2022ocm}.

The density operator $n$ has  mass-dimension $3$. The mass dimension as well as the scaling dimension of the bosonic degrees of freedom $\phi(t,\boldsymbol{x},\theta)$ is zero. Note that  $\phi$ always comes with one $\nabla_x/p_F$.

{ Let's count how many times $p_F$ appears. Diagram (A) in Fig.~\ref{fig:omega meson propagator in NB} has a vertex coming from Eq.~(\ref{eq:bosonized lagrangian2}) $\sim p_F^3/m_N^\sigma$, with $m_N^\sigma\sim p_F$ at $n_0$, the diagram (A) is $\sim p_F^2$. The diagram (B) has two vertexes $\sim p_F^3/m_N^\sigma$ and a internal $\phi$ line $\sim p_F^{-2}$, so  (B) is $\sim p_F^2$.

For loop diagrams in the bosonic theory, every internal $\omega$ meson line contributes $p_F^{-2}$ because $m^\ast_\omega \sim p_F$. Every loop integral contributes $p_F^4$. Diagram (C) has two vertices $\sim p_F^2/m_N^\sigma$, a loop $\sim p_F^4$, two internal lines $\sim p_F^{-2}$.  There is a caveat here as will be elaborated in Fig.~\ref{fig:FL};  a $\phi$ line carrying large momentum $\boldsymbol{k}$ is suppressed by a factor $\Lambda/p_F$ \cite{Shankar:1993pf}. So (C) is $\sim p_F$. Other diagrams are shown in Fig.~\ref{fig:omega meson propagator in NB}, with D meaning the diagram is $\sim p_F^D$. }
\begin{figure*}[!ht]
    \centering
    \includegraphics[width=0.75\textwidth]{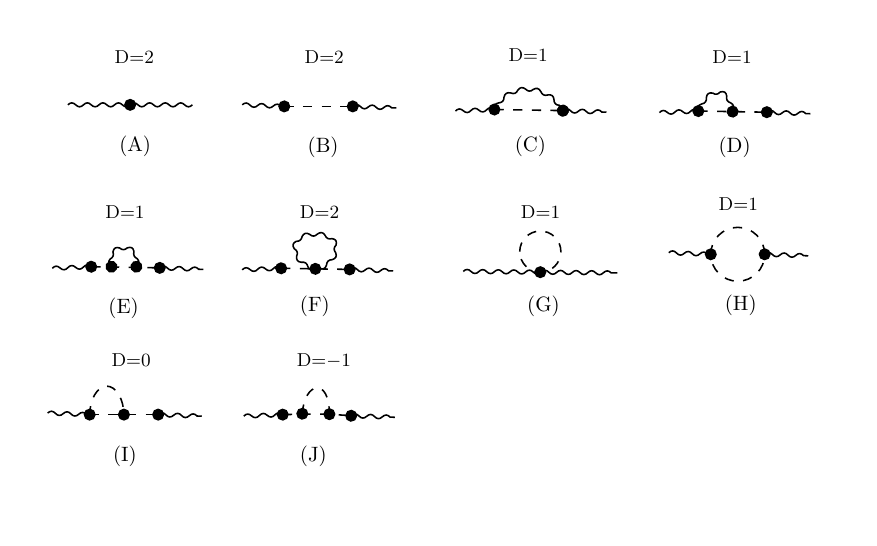}
    \caption{Corrections to the $\omega$ meson propagator in nonlinear bosonized theory}
	\label{fig:omega meson propagator in NB}
\end{figure*}

The diagrams (A)-(B) are  the leading-order corrections to the free $\omega$ meson propagator. There are many diagrams in the next-to the leading orders. However, most diagrams do not  contribute in the limit that the external momentum $\boldsymbol{q}/q_0\rightarrow0$. The next-to-leading order contributions to the meson propagator should be the $D=1$ diagrams, which contain (C),(D),(E),(G),(H). This power counting is valid for one-loop graphs. For higher loop diagrams, however, there are more suppressions coming from phase space arguments with which we are not concerned in this work.

\section{\label{app:2}Corrections to the $\omega$ meson propagator}
\subsection{\label{app:2.1}Leading-order corrections}

Amputating the $\omega$ line,  (A) in Fig.~(\ref{fig:omega meson propagator in NB}) is simply $n/m_N^\sigma$ with $n=\nu p_F^3/6\pi^2$. And the (B) is given by
\begin{widetext}
    \begin{equation}
    \begin{aligned}
        (B) &= -i\frac{p_F^4}{(2\pi)^6}\int \frac{d\boldsymbol{p}}{(2\pi)^3}\delta(|\boldsymbol{p}|-p_F)\boldsymbol{n}_\theta\cdot i\boldsymbol{q}\int \frac{d\boldsymbol{p'}}{(2\pi)^3}\delta(|\boldsymbol{p'}|-p_F)\boldsymbol{n}_\theta'\cdot i(-\boldsymbol{q})\left\langle \phi\phi'\right\rangle\frac{p^ip^{i'}}{m^{\sigma2}_N} \\
        &=\frac{p_F^2}{(2\pi)^3}\frac{\nu p_F^2}{3m^{\sigma2}_N}\int d^2\theta\frac{\boldsymbol{n}_\theta\cdot\boldsymbol{q}}{q_0-\frac{p_F}{m_N^\sigma}\boldsymbol{n}_\theta\cdot\boldsymbol{q}} \\
        &=-\frac{\nu p_F^3}{6\pi^2m^{\sigma}_N}\Bigg(1+\frac{m^{\sigma}_N|q_0|}{2p_F|\boldsymbol{q}|}\log\frac{m_N^\sigma|q_0|-p_F|\boldsymbol{q}|}{m_N^\sigma|q_0|+p_F|\boldsymbol{q}|}\Bigg)    
    \end{aligned}   \label{eq:B}
\end{equation}
\end{widetext}
where $q_\mu=(q_0,\boldsymbol{q})$ is external momentum. The sum of (A) and (B) reproduces Eq.~(\ref{eq:polarization tensor}).

\subsection{\label{app:2.2}Next-to-leading order corrections}

To calculate the diagram (C) in Fig.~\ref{fig:omega meson propagator in NB}, we have to divide the loop momentum  into two parts, one from $0$ to $\Lambda$, denoted as part \uppercase\expandafter{\romannumeral1}, and the other from $\Lambda$ to $2p_F$, denoted as part \uppercase\expandafter{\romannumeral2}. While the part \uppercase\expandafter{\romannumeral1} is not suppressed by phase space argument, the part II is suppressed by a factor $\Lambda/p_F$~\cite{Shankar:1993pf}. However the part \uppercase\expandafter{\romannumeral1} is  suppressed by $(\Lambda/p_F)^3$ compared to the part \uppercase\expandafter{\romannumeral2} due to the integration region, so the region \uppercase\expandafter{\romannumeral2} should give the dominant contribution
\begin{widetext}
    \begin{equation}(C)=\int_{\uppercase\expandafter{\romannumeral2}}\frac{d^3\boldsymbol{k}dk_0}{(2\pi)^4}\frac{1}{m^{\sigma2}_N}\frac{-ig_\omega^2}{(q_\mu-k_\mu)^2-m_\omega^{\ast2}+i\epsilon}\frac{\nu p_F^2}{2(2\pi)^3}\int d^2\theta\frac{\boldsymbol{n}_\theta\cdot\boldsymbol{k}}{k_0-\frac{p_F}{m_N^\sigma} \boldsymbol{n}_{\boldsymbol{\theta}}\cdot\boldsymbol{k}+i\epsilon}   \label{eq:C}
    \end{equation}
\end{widetext}
where $k_\mu=(k_0,\boldsymbol{k})$ is the loop momentum. The reason why the part \uppercase\expandafter{\romannumeral2} is suppressed by $\Lambda/p_F$ is shown in Fig.~\ref{fig:FL}, given in the fermionic description
\begin{figure}[!ht]
    \centering
    \includegraphics[width=0.5\textwidth]{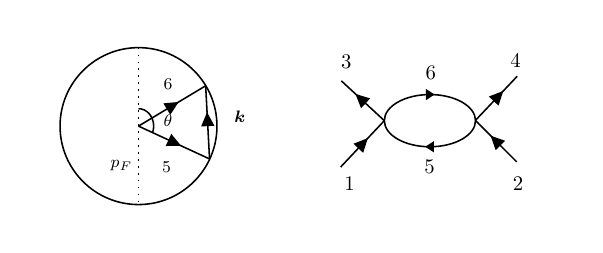}
	\caption{The phase space restriction when momentum transfer $\boldsymbol{k}$ is large}
	\label{fig:FL}
\end{figure}

When we take $\Lambda\rightarrow0$, then all fermions 1-6 lie at the Fermi surface. Suppose the difference between $\boldsymbol{k_3}$ and $\boldsymbol{k_1}$ is $\boldsymbol{k}$. When the momentum transfer $|\boldsymbol{k}|$ is within $\Lambda$, i.e. $|\boldsymbol{k}|\rightarrow0$, the loop momentum $\boldsymbol{k_5}$ runs freely at the Fermi surface. However, when $\boldsymbol{k}$ is large like the one in Fig.\ref{fig:FL}, $\boldsymbol{k_5}$ is restricted in a specific point at the Fermi surface corresponding to the polar angle $\theta$. When $\Lambda$ is not exactly zero, the specific point will expand to a region of angle $\Lambda/p_F$. This argument is made in 2D. In 3D, there is an extra azimuthal angle that allows the fermions 5 and 6 to run freely. 

In the bosonic theory, the second integral in Eq.~(\ref{eq:C}) is easy to work out. When the angle $\theta$ is in a fixed region rather than the whole Fermi surface, the integrand is a quantity dependent of $|\boldsymbol{k}|$ and independent of angle. The integral will be replaced by the integrand times the angle region.  After some manipulation,  $\boldsymbol{n}_\theta\cdot\boldsymbol{k}=-\boldsymbol{k}^2/2p_F$. So (C) becomes
\begin{widetext}
\begin{equation}
     \begin{aligned}
    (C)=&i\frac{\nu\Lambda p_F}{2(2\pi)^2m_N^{\sigma2}}\int_{\uppercase\expandafter{\romannumeral2}}\frac{d\boldsymbol{k}dk_0}{(2\pi)^4}\frac{g_\omega^2}{(q_\mu-k_\mu)^2-m_\omega^{\ast2}+i\epsilon}\frac{\frac{\boldsymbol{k}^2}{2p_F}}{k_0+\frac{\boldsymbol{k}^2}{2m_N^\sigma}+i\epsilon}  \\=& i\frac{\nu\Lambda g_\omega^2}{16\pi^2m_N^{\sigma2}}\int_{\uppercase\expandafter{\romannumeral2}}\frac{d\boldsymbol{k}dk_0}{(2\pi)^4}\frac{\boldsymbol{k}^2}{\left[ k_0-(q_0+\sqrt{(\boldsymbol{q}-\boldsymbol{k})^2+m_\omega^{\ast2}})+i\epsilon\right]\left[ k_0-(q_0-\sqrt{(\boldsymbol{q}-\boldsymbol{k})^2+m_\omega^{\ast2}})-i\epsilon\right](k_0+\frac{\boldsymbol{k}^2}{2m^\sigma_N}+i\epsilon) }  \\
    =&-\frac{\nu\Lambda g_\omega^2}{64\pi^4m_N^{\sigma2}}\int_\Lambda^{2p_F} dk\frac{k^4}{k^2+m_\omega^{\ast2}-\frac{k^2\sqrt{k^2+m_\omega^{\ast2}}}{2m^\sigma_N}}   \label{eq:C2}
    \end{aligned}
\end{equation}
\end{widetext}

The nonlinear bosonized theory has some similarity to nonlinear sigma model. For this reason, the diagrams (D-F) vanish in the limit the momentum transfer goes to zero. It can be shown that the diagrams   (G) and (H) also vanish in the limit. Therefore (C) is the only diagram that needs to be calculated.

In the limit $|\boldsymbol{q}|/q_0\rightarrow0, q_0\rightarrow0$, the nonlinear structure of the Fermi surface does not affect the next leading order corrections to the $\omega$ meson propagator because (C) does not contain the vertex that connects two dashed lines. The diagrams (D) and (E) can actually be seen as loop corrections to the $\phi$-$\omega$ vertex. If not amputated, the $\phi$-$\omega$ vertex in the nonlinear bosonized theory corresponds to the $\omega$-$N$-$N$ vertex in~\cite{Paeng:2013xya}. So the (D) and (E) vanishing in the limit $|\boldsymbol{q}|/q_0\rightarrow0$ should be seen as a result of unscaling of $g_{\omega NN}$ in one fermion-loop order.

$$$$

\end{document}